\documentclass[prb,aps,twocolumn,superscriptaddress,showpacs,10pt]{revtex4-1}
\pdfpageattr {/Group << /S /Transparency /I true /CS /DeviceRGB>>}
\usepackage{hyperref}
\hypersetup{%
pdftitle={Measurement of exciton interactions using electrostatic lattices},%
pdfauthor={M. Remeika et al.},%
pdfpagemode={UseNone},%
pdfstartview={FitH},%
breaklinks=true,%
citecolor=blue,%
colorlinks=true,%
linkcolor=blue,%
urlcolor=blue}

\usepackage{amsmath,amssymb,bm,graphicx}

\newcommand{\unit}[1]{\,\mathrm{#1}} 
\newcommand{\sub}[1]{$_\text{#1}$} 

\begin{document}

\title{Measurement of exciton correlations using electrostatic lattices}

\author{M. Remeika}
\author{J.~R. Leonard}
\email{jleonard@physics.ucsd.edu}
\author{C.~J. Dorow}
\author{M.~M. Fogler}
\author{L.~V. Butov}
\affiliation{Department of Physics, University of California at San Diego, La Jolla, CA 92093-0319, USA}

\author{M. Hanson}
\author{A.~C. Gossard}
\affiliation{Materials Department, University of California at Santa Barbara, Santa Barbara, CA 93106-5050, USA}

\begin{abstract}

We present a method for determining correlations in a gas of indirect excitons in a semiconductor quantum well structure. The method involves subjecting the excitons to a periodic electrostatic potential that causes modulations of the exciton density and photoluminescence (PL). Experimentally measured amplitudes of energy and intensity modulations of exciton PL serve as an input to a theoretical estimate of the exciton correlation parameter and temperature. We also present a proof-of-principle demonstration of the method for determining the correlation parameter and discuss how its accuracy can be improved.

\end{abstract}

\date{\today}

\maketitle

\section{Introduction}
\label{sec:Introduction}

\begin{figure}[b]
\begin{center}
\includegraphics[width=3.0in]{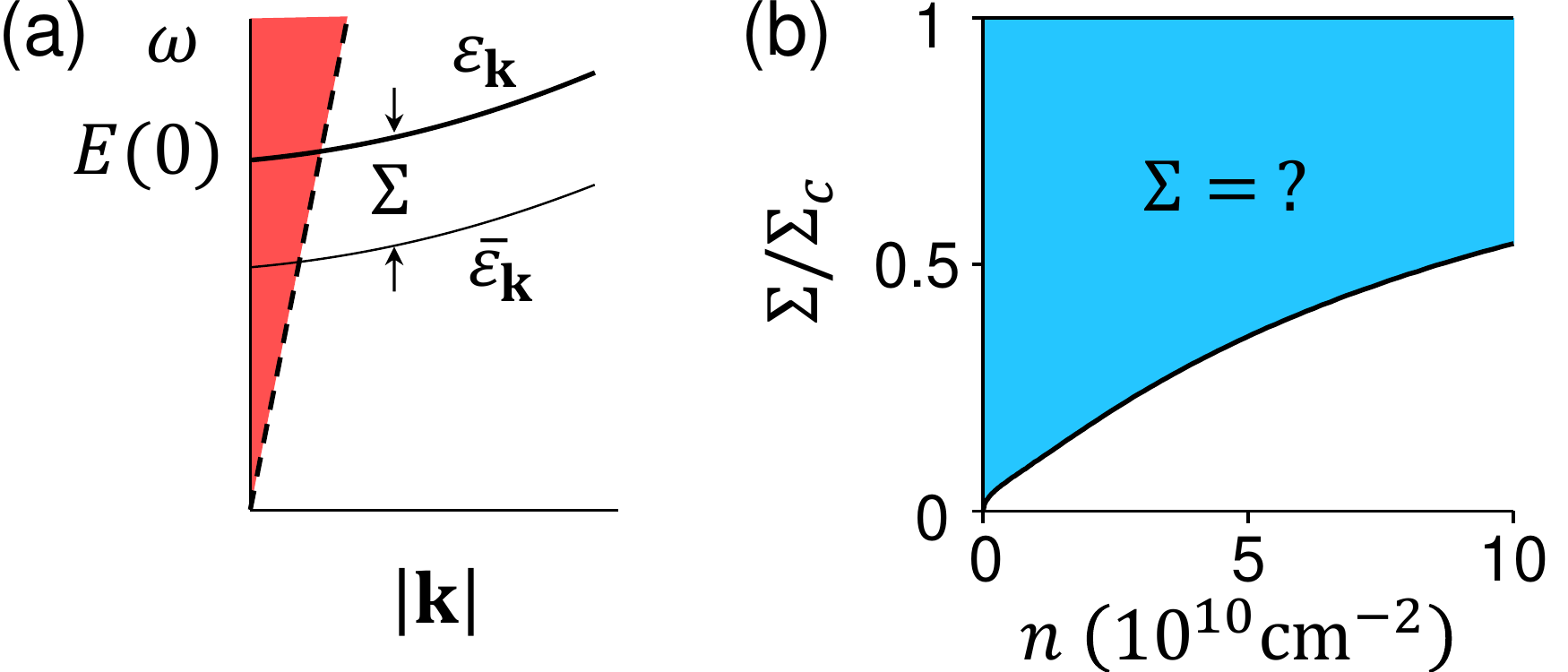}
\end{center}
\caption{(Color online) (a) Bare and renormalized exciton dispersions (thin and thick solid lines, respectively; both are schematic). The dashed line is the light dispersion; the shaded region to the left of it is the radiative zone. (b) The range of values for exciton self-energy $\Sigma$ (shaded). The upper solid line is the Hartree (capacitance) self-energy $\Sigma_c$ to which $\Sigma$ is normalized in this plot. The lower solid line is a lower bound estimated from the self-energy of an exciton crystal with parameters $d = 1.2\, a_e$, $a_e = 10\unit{nm}$, and $m_h = 2\, m_e$ suitable for GaAs CQW studied in this work. A hydrogenic exciton wavefunction $\phi(r_\mathrm{eh}) = \exp(-r_\mathrm{eh} /\, a)$ was assumed with $a = 2.9\, a_e$.
}
\label{fig:cartoons}
\end{figure}

Indirect excitons (IXs) in coupled quantum well structures (CQW)~\cite{Lozovik1976nms, Fukuzawa1990pcl} is a model system for exploring diverse physical phenomena including pattern formation,~\cite{Butov2002mos} spontaneous coherence and condensation,~\cite{Yang2006clc, Fogler2008esr, High2012sci, High2012cei, Alloing2014ebe} transport,~\cite{Hagn1995eft, Butov1998atl, Larionov2002csi, Gartner2006dml, Ivanov2006oir, Hammack2007kie, Hammack2009kir, Lazic2010etm, Alloing2012ndi, Lazic2014sir, Fedichkin2015tde, Kuznetsova2015tie} spin transport and spin textures,~\cite{High2013scc} and localization-delocalization transitions.~\cite{Remeika2009ldt, Winbow2011ece, Remeika2012tde} The CQW consists of a pair of parallel quantum wells separated by a narrow tunneling barrier and the IX is a bound state of an electron and a hole confined in the opposite quantum wells. Interactions play a key role in the physics of IX systems. In the first approximation, the exciton-exciton interaction potential is given by
\begin{equation}
v(r) = \frac{2 e^2}{\kappa}
       \left(\frac{1}{r} - \frac{1}{\sqrt{r^2 + d^2}}\right)
       ,
\label{eqn:V}
\end{equation}
where $d$ is the center-to-center separation of the two wells in the CQW and $\kappa$ is the dielectric constant of the semiconductor. At $r \gg d$ this potential has the form of the dipole-dipole repulsion, $v(r) \simeq e^2 d^2 / \kappa r^3$.
A hallmark of interacting IXs is the increase of its photoluminescence (PL) energy $E$ with density $n$, which has been known since early spectroscopic studies of CQWs.~\cite{Butov1994cie, Butov1999mos} At small $n$, where the interactions are negligible, this energy approaches the creation energy $E(0)$ of a single IX. As $n$ increases, a pronounced PL blue shift
\begin{equation}
 \Delta E(n) = E(n) - E(0) > 0
\label{eqn:DeltaE}
\end{equation}
develops. The physical origin of this energy shift is the renormalization of the IX dispersion. The bare dispersion is given by
\begin{equation}
\bar\varepsilon_\mathbf{k} = E(0)
 + \frac{\hbar^2 \mathbf{k}^2}{2 m_x}\,,
\label{eqn:varepsilon_k}
\end{equation}
where $\mathbf{k}$ is the exciton momentum, see Fig.~\ref{fig:cartoons}(a). (In this paper the bare quantities are denoted with bars.) The renormalized dispersion $\varepsilon_\mathbf{k}$ is the solution of the equation $\varepsilon_\mathbf{k} = \bar\varepsilon_\mathbf{k} + \Sigma^\prime(\varepsilon_\mathbf{k})$, where $\Sigma^\prime$ is a real part of the self-energy
\begin{align}
\Sigma(\varepsilon) = \Sigma^\prime(\varepsilon) +
i\/ \Sigma^{\prime\prime}(\varepsilon)\,.
\label{eqn:Delta_from_Sigma}
\end{align}

\begin{figure}[t]
\begin{center}
\includegraphics[width=3.0in]{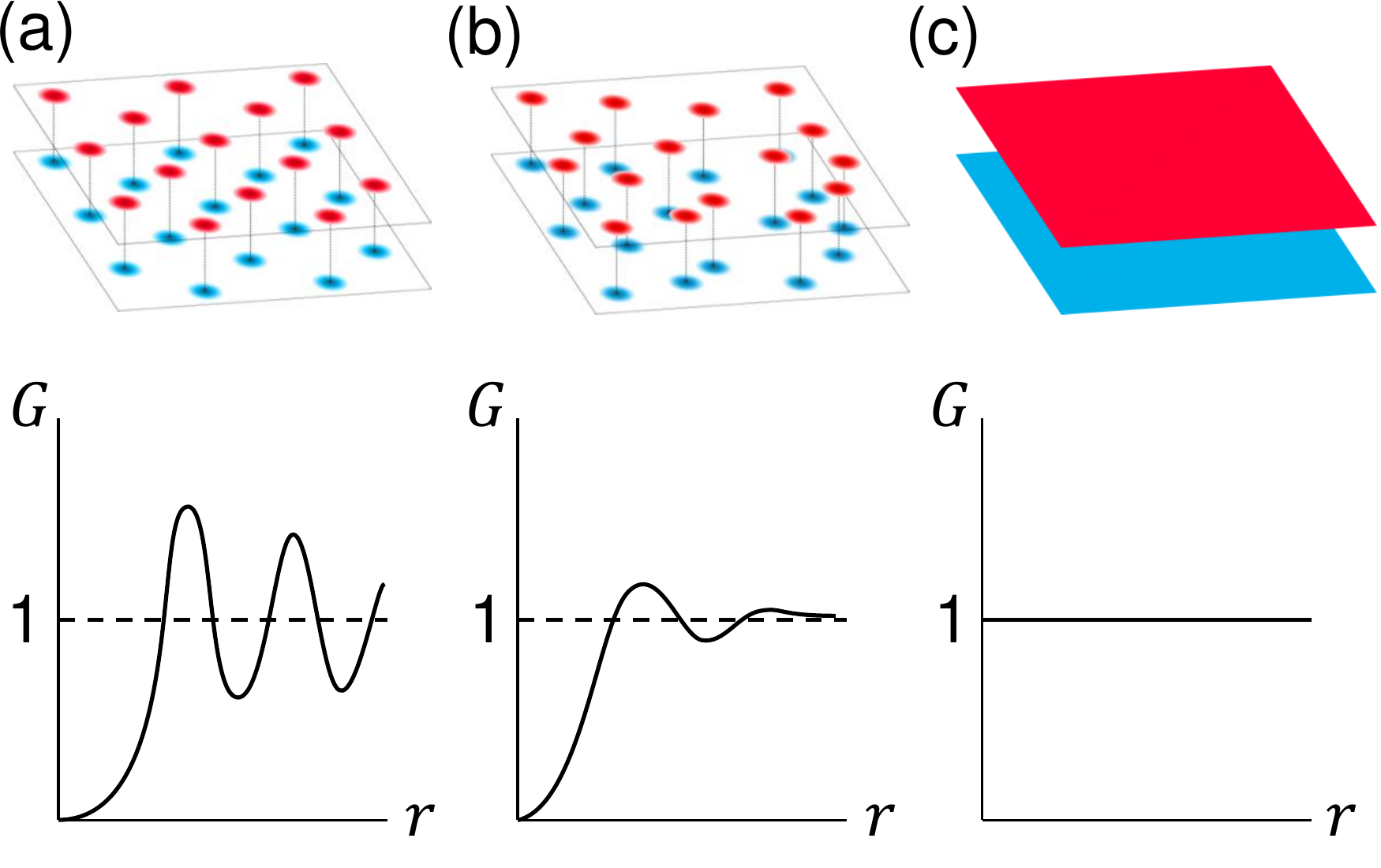}
\end{center}
\caption{(Color online) Illustrations of indirect exciton phases: (a) crystal (b) liquid with short-range correlations (c) hypothetical uncorrelated liquid or gas. The blue (red) patches represent electrons (holes) confined in separate 2D layers. The bottom row shows the schematic plots of two-body correlation functions $G({r}) = n^{-2} \langle n(\mathbf{r}) n(0)\rangle - \delta(\mathbf{r})$ for each of the phases.
}
\label{fig:cartoons2}
\end{figure}

The simplest theoretical model of the self-energy is the Hartree approximation. It assumes that $\Sigma$ is equal to the energy-independent, real constant
\begin{align}
\Sigma_c = n \tilde{v},
\quad
\tilde{v} \equiv \int v({r}) d^2 r
 = \frac{4\pi e^2 d}{\kappa}\,,
\label{eqn:t_0}
\end{align}
so that the PL energy shifts by the same amount:
\begin{equation}
\Delta E = \Sigma_c
         = \frac{4\pi e^2}{\kappa}\, n d\,.
\label{eqn:Sigma_naive}
\end{equation}
Equation~\eqref{eqn:Sigma_naive} is colloquially known as the ``capacitor'' formula~\cite{Ivanov2002qdd} because it is similar to the expression for the voltage on a parallel-plate capacitor with surface charge density $\pm e n$ on the plates. The capacitor formula provides a qualitative explanation for the observed monotonic increase of $\Delta E$ with photoexcitation power. However, analytical theory beyond the Hartree approximation~\cite{Yoshioka1990dqw, Zhu1995ecs, Lozovik1997pti, Ben-Tabou_de-Leon2003mtb, Zimmermann2007eii, Schindler2008aee, Laikhtman2009exc, Ivanov2010cpr} and Monte-Carlo calculations~\cite{DePalo2002eci, Maezono2013eab} suggest that the capacitor formula significantly overestimates $\Delta E(n)$. The same conclusion follows from the analysis of the small-$n$ and the large-$d$ limits, where IX should form, respectively, a correlated liquid and a crystal, see Figs.~\ref{fig:cartoons2}(a) and (b).
In such phases the IXs avoid each other, which lowers their interaction energy per particle as well as $\Sigma^{\prime}$
compared with the uncorrelated gas-like state assumed in the Hartree approximation, Fig.~\ref{fig:cartoons2}(c).

For comparison of various theories with experiment and with one another, we define the dimensionless correlation parameter~\cite{Remeika2009ldt}
\begin{equation}
\gamma = \bar\nu_1 \Delta E /\, n\,,
\label{eqn:gamma}
\end{equation}
where
\begin{equation}
\bar\nu_1 = {m_x}\,/\,{2\pi \hbar^2}
\label{eqn:nu_1}
\end{equation}
is the bare exciton density of states (DOS) per spin and $m_x = m_e + m_h$ [Eq.~\eqref{eqn:varepsilon_k}], $m_e$, and $m_h$, are the effective masses of, respectively, excitons, electrons, and holes. The capacitor formula~\eqref{eqn:Sigma_naive} predicts the density-independent correlation parameter
\begin{equation}
\gamma_c = \bar\nu_1 \tilde{v}
         = \frac{2 d}{a_e}\,\frac{m_e + m_h}{m_e}\,,
\label{eqn:gamma_plate}
\end{equation}
which is about $\gamma_c \approx 7$ for the GaAs CQW structures studied in Ref.~\onlinecite{Remeika2009ldt} and the present work. Here $a_e = \hbar^2 \kappa / (m_e e^2)$ is the electron Bohr radius. The interaction part of the self-energy of a classical crystal gives a lower bound on $\Sigma^{\prime}$ and thus $\gamma$. However, this leaves one with a large uncertainty, see Fig.~\ref{fig:cartoons}(b). A simple remedy for the inaccuracy of the Hartree approximation is to replace the coefficient $\tilde{v}$ in Eq.~\eqref{eqn:t_0} with a smaller number $t$. This yields $\Delta E = \Sigma^{\prime}= n t$ and
\begin{equation}
\gamma = \bar\nu_1 t\,.
\label{eqn:gamma_from_t}
\end{equation}
The problem remains how to reliably evaluate the phenomenological parameter $\gamma$. The solution of this problem can provide both the test of available theories and a convenient density calibration tool in experiments.

\begin{figure}[b]
\centering
\includegraphics[width=3.2in]{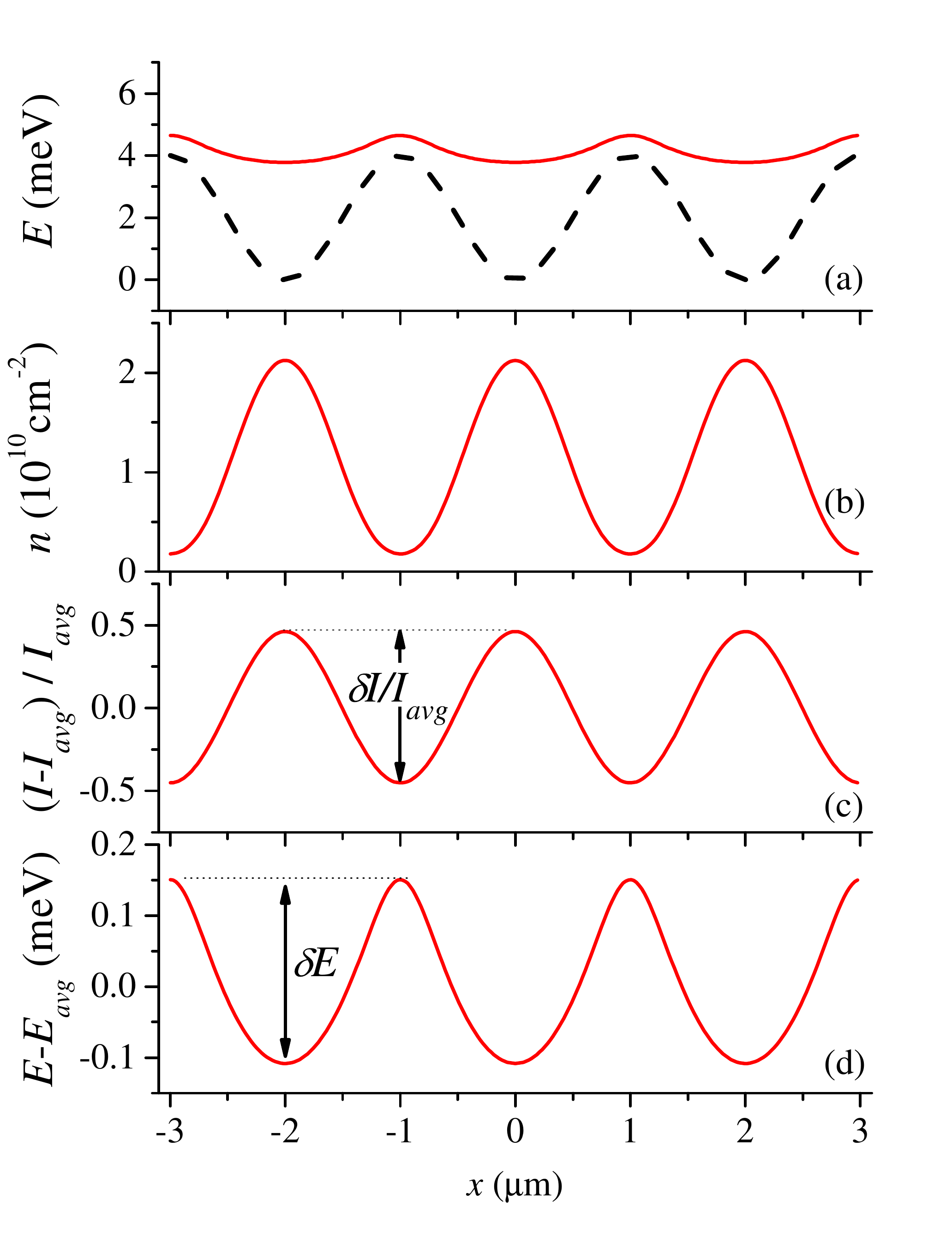}
\caption{(Color online) Simulated PL energy and intensity of indirect excitons in a lattice potential. (a,b) Lattice potential [dashed line in (a)], PL energy [solid line in (a)], and exciton density (b) for $\zeta = 3\unit{meV}$ (measured with respect to $\bar\varepsilon_0$). (c,d) PL intensity (c) and energy (d) under the same conditions for $\mathrm{NA} = 0.4$. The depth of the lattice $U_l = 4\unit{meV}$, temperature $T = 4.2\unit{K}$, correlation parameter $\gamma = 7$.}
\label{fig:simulated_profiles}
\end{figure}

\begin{figure}
\begin{center}
\includegraphics[width=2.0in]{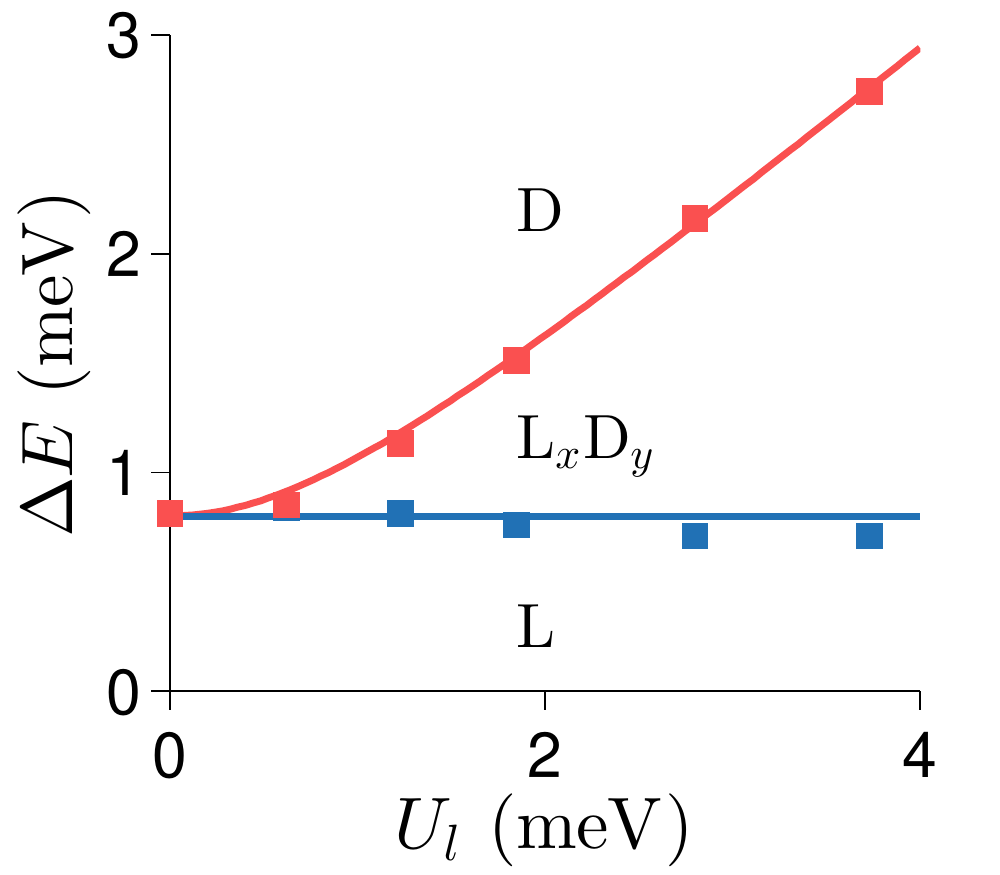}
\end{center}
\caption{(Color online) Localization-delocalization crossovers. Indirect excitons are localized in region $\mathrm{L}$, localized along $x$ but delocalized along $y$ in region $\mathrm{L}_{x}\mathrm{D}_{y}$, and delocalized in both directions in region $\mathrm{D}$. The last of these is suitable for measuring $\gamma$. The squares are data from our earlier work.~\cite{Remeika2009ldt} The lines are empirical fits $\Delta E = \left(0.8^2 + 0.5 U_l^2\right)^{1/2}$ and $\Delta E = 0.8$ (in units of $\mathrm{meV}$).
}
\label{fig:phases}
\end{figure}

In our previous work~\cite{Remeika2009ldt} we outlined a method that allows one to estimate $\gamma$ experimentally. The key idea of the method is that the same correlation parameter determines both the PL shift and the ability of IXs to screen an external potential perturbation. In other words, the screening efficiency of the IX gas is related to $\gamma$. In the experimental part of this work, we infer the screening properties of the IX gas from variation of its PL energy and intensity as a function of coordinates. To do so we employ a periodic external potential
\begin{equation}
U(x) = \frac12 U_l (1 - \cos q x)
\label{eqn:E_0}
\end{equation}
created with the help of interdigitated gate electrodes. The ``depth'' $U_l$ of this potential is controlled by the gate voltage. As the IXs attempt to screen the potential, their density becomes periodically modulated (Fig.~\ref{fig:simulated_profiles}). In our previous work,~\cite{Remeika2009ldt} we measured modulation of the IX PL intensity and energy in such a device. Analyzing such experiments using a simplified theoretical model, we arrived at a rough estimate $\gamma \sim 2$, which is about one-third of $\gamma_c$. This analysis requires large enough $U_l > \Gamma \sim 0.8\unit{meV}$ to ensure that the lattice potential dominates over the random potential of disorder. In addition, sufficient IX density producing $\Delta E$ higher than $U_l$ and $\Gamma$ is required in order to overcome localizations effects of both these potentials (Fig.~\ref{fig:phases}) as detailed in Sec.~\ref{sec:Model}. In the present work we expand these earlier efforts into a fully fledged method for determining correlations in a gas of IXs from the measured modulations of exciton PL in a periodic electrostatic potential. Our refined model takes into account optical resolution effects neglected in Ref.~\onlinecite{Remeika2009ldt}.

It is worthwhile to note that a conceptually similar technique has been previously employed for cold atoms in magnetic traps. The equilibrium density profile of the atomic gas in the trap can be understood as a result of screening the trapping potential by the atoms. Furthermore, it has been shown that such density profiles can be measured optically and used to determine the equation of state of the gas.~\cite{Shin2008des, Bloch2012qsu} By analogy, one can think of our external potential $U(x)$ as a periodic array of individual traps. This geometry offers certain advantages compared to single traps. One is the redundancy of the experimental data, which reduces stochastic errors of the measurements using uniform arrays. Another advantage is the simplicity of modeling the optical resolution effects in a periodic geometry.

The remainder of the paper is organized as follows. In Sec.~\ref{sec:Model} we define the model and derive the main equations of the theory. In Sec.~\ref{sec:Experiments} we present a proof-of-principle demonstration of the method for determining the correlation parameter. In Sec.~\ref{sec:Discussion} we discuss how to improve the accuracy of the method.

\section{Theoretical model}
\label{sec:Model}

Our objective is to describe PL characteristics of an exciton gas subject to a periodic external perturbation, Eq.~\eqref{eqn:E_0}, which makes the exciton density modulated. However, it is instructive to start with the case $U_l = 0$ where the exciton gas is uniform and the in-plane momentum $\mathbf{k}$ is still a good quantum number.

\subsection{Uniform exciton gas}
\label{sub:uniform}

The PL intensity spectrum $\tilde{I}(\omega)$ in our model is given by the formula
\begin{equation}
\tilde{I}(\omega) = B f(\omega - \mu) A_{0}(\omega)\,,
\label{eqn:I_II}
\end{equation}
where $B$ is a constant related to the exciton-photon matrix elements,
\begin{equation}
f(\varepsilon) = \frac{1}{e^{\beta \varepsilon} - 1}
\label{eqn:f_Bose}
\end{equation}
is the Bose-Einstein distribution function, $\beta = 1 / T$ is the inverse temperature, and $\mu$ is the exciton chemical potential. The optical density (OD) $A_{0}(\omega)$ is the $\mathbf{k} = 0$ part of the spectral function
\begin{equation}
A_{\mathbf{k}}(\omega)
= -2\,\mathrm{Im}\,
\frac{1}
     {\omega - \bar\varepsilon_{\mathbf{k}} - \Sigma_{\mathbf{k}}}\,.
\label{eqn:A_from_Sigma}
\end{equation}
Derivation of Eq.~\eqref{eqn:I_II} can be found in earlier works on the subject.~\cite{Feldmann1987ldr, Hanamura1998rrd, Andreani1991rlf, Citrin1993rle, Runge2003, Zimmermann2003} When interactions and disorder are present in the system, the functional form of OD may be complicated and qualitatively different in the canonical phases of bosonic matter: normal fluid, superfluid, and Bose glass.~\cite{Giamarchi1988ali, Fisher1989bls} We focus on the intermediate density range,
\begin{equation}
n_d \gtrsim n \gg \frac{T}{t} = \frac{n_d}{g \gamma}\,,
\label{eqn:n_domain}
\end{equation}
in which excitons are expected to form the normal fluid. Parameter $g$ in Eq.~\eqref{eqn:n_domain} is the number of exciton spin flavors ($g = 4$ in GaAs) and
\begin{equation}
n_d = g \bar\nu_1 T
\label{eqn:n_d}
\end{equation}
is the density scale corresponding to the onset of quantum degeneracy. We assume that the interactions and disorder shift the exciton energies by the $\mathbf{k}$-independent amount $\Sigma^{\prime} = n t$, so that
\begin{equation}
\varepsilon_\mathbf{k} = \bar\varepsilon_\mathbf{k} + n t\,,
\label{eqn:varepsilon_k_renorm}
\end{equation}
and that the exciton scattering rate $|\Sigma^{\prime \prime}|$ is small compared to the temperature $T$. Under these simplifying assumptions, the OD as a function of $\omega$ can be approximated by a single sharp peak centered at the renormalized exciton band edge $\varepsilon_0 = \bar\varepsilon_0 + n t$.

We restrict our analysis to the two most basic characteristics of the PL spectrum, the total intensity and the average energy,
which we define using the moments of the lineshape:
\begin{equation}
I^{(m)} \equiv \int\limits_{0}^{\infty} \omega^m
      \tilde{I}(\omega) \frac{d \omega}{2\pi}\,.
\label{eqn:I^m}
\end{equation}
Specifically, the PL intensity is the zeroth moment and the energy is the ratio of the first and zeroth moments:
\begin{equation}
I = I^{(0)}\,,
\quad
E \equiv {I^{(1)}} / {I^{(0)}}\,.
\label{eqn:E_def}
\end{equation}
It is easy to find the relations among the key quantities in this model:
\begin{align}
n &= -n_d \ln
\left[1 - e^{\beta(\mu - \varepsilon_0)}\right],
\quad
\mu = \bar\mu + n t\,,
\label{eqn:n_from_mu}\\
I &= \frac{B}{e^{\beta(\varepsilon_0 - \mu)} - 1}
= B \left(e^{n / n_d} - 1\right),
\label{eqn:averageI}\\
E &= \varepsilon_0 = \bar\varepsilon_0 + n t\,.
\label{eqn:averageE}
\end{align}

\begin{figure}[t]
\centering
\includegraphics[width=1.8in]{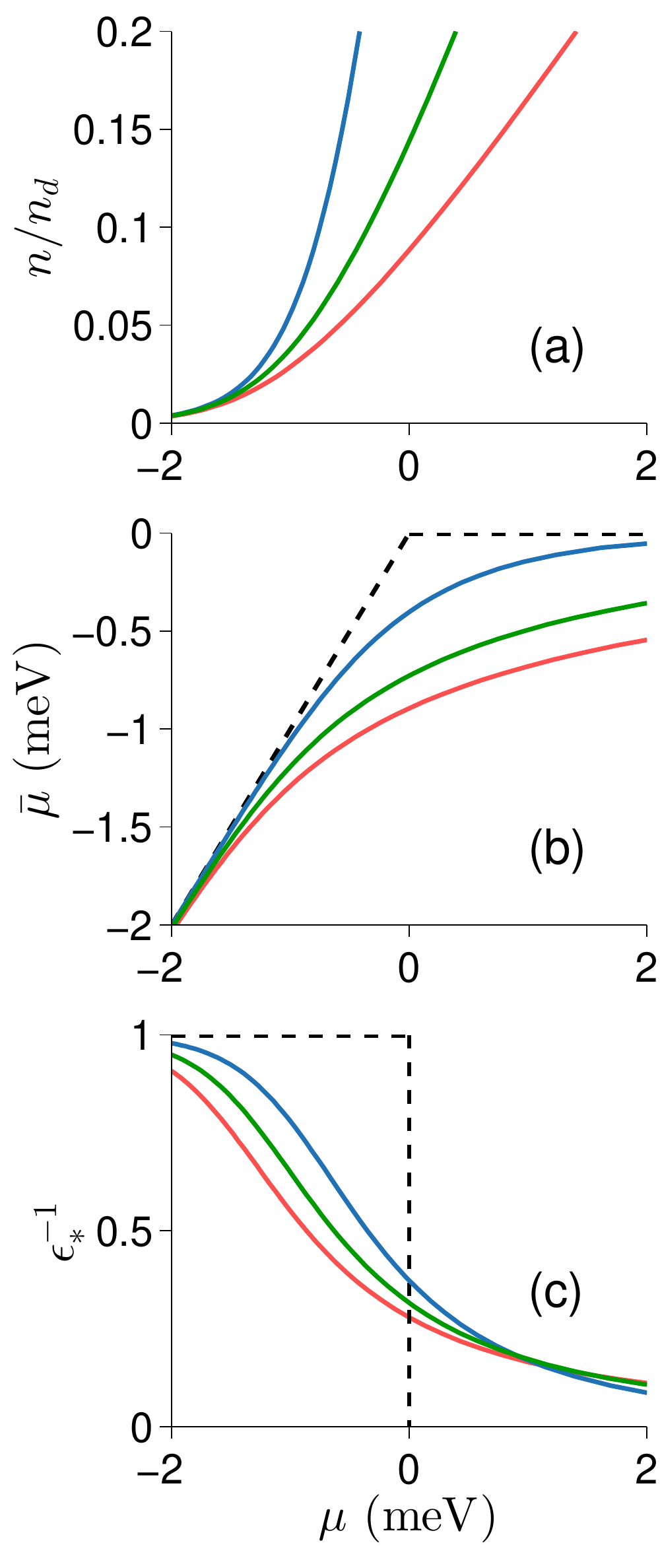}
\caption{(Color online) Exciton density, chemical potentials, and effective dielectric constant. (a) Density $n$ as a function of the (renormalized) chemical potential $\mu$ for (left to right) $\gamma = 0.7$, $3.5$, $7.0$. (b) The bare chemical potential as a function of the renormalized one for the same $\gamma$ (solid lines). (c) Inverse $\epsilon_*$ as a function of $\mu$. The dashed lines in (b, c) depict the $\gamma \to 0$ limit. $T = 4.2\unit{K}$ for all the curves.
}
\label{fig:mu_and_n}
\end{figure}

\noindent{}Equation~\eqref{eqn:n_from_mu} follows from
\begin{equation}
n = g \int\limits_{-\infty}^{\infty} {\nu}_1(\omega)
  f(\omega - \mu) d\omega
\label{eqn:n_from_nu_bar}
\end{equation}
where
$
{\nu}_1(\omega) =
 \bar\nu_1\, \theta(\omega - \varepsilon_0)
$
is the renormalized DOS. [By virtue of Eq.~\eqref{eqn:varepsilon_k_renorm}, the renormalization is simply a shift of the argument by $n t$.] The first equality in Eq.~\eqref{eqn:averageI} comes from Eq.~\eqref{eqn:I_II} and the normalization condition
\begin{equation}
\int\limits_{-\infty}^{\infty} A_0(\omega)
 \frac{d \omega}{2\pi}
 = 1
\label{eqn:A_0_normalization}
\end{equation}
for the OD. Equation~\eqref{eqn:averageE} establishes the correspondence
\begin{equation}
E(0) = \bar\varepsilon_0\,,\quad \Delta E(n) = n t
\end{equation}
with the notations introduced in Sec.~\ref{sec:Introduction}. While the density dependence of $E$ is linear, that of $\mu$ is not, see Fig.~\ref{fig:mu_and_n}(a). The nonlinear relation between $\mu$ and $\bar\mu$ at several different $\gamma$ is illustrated by Fig.~\ref{fig:mu_and_n}(b). This relation characterizes screening properties of the exciton fluid discussed next in Sec.~\ref{sub:modulated}.

\subsection{Periodically modulated exciton gas}
\label{sub:modulated}

\begin{figure}[b]
\centering
\includegraphics[width=3.2in]{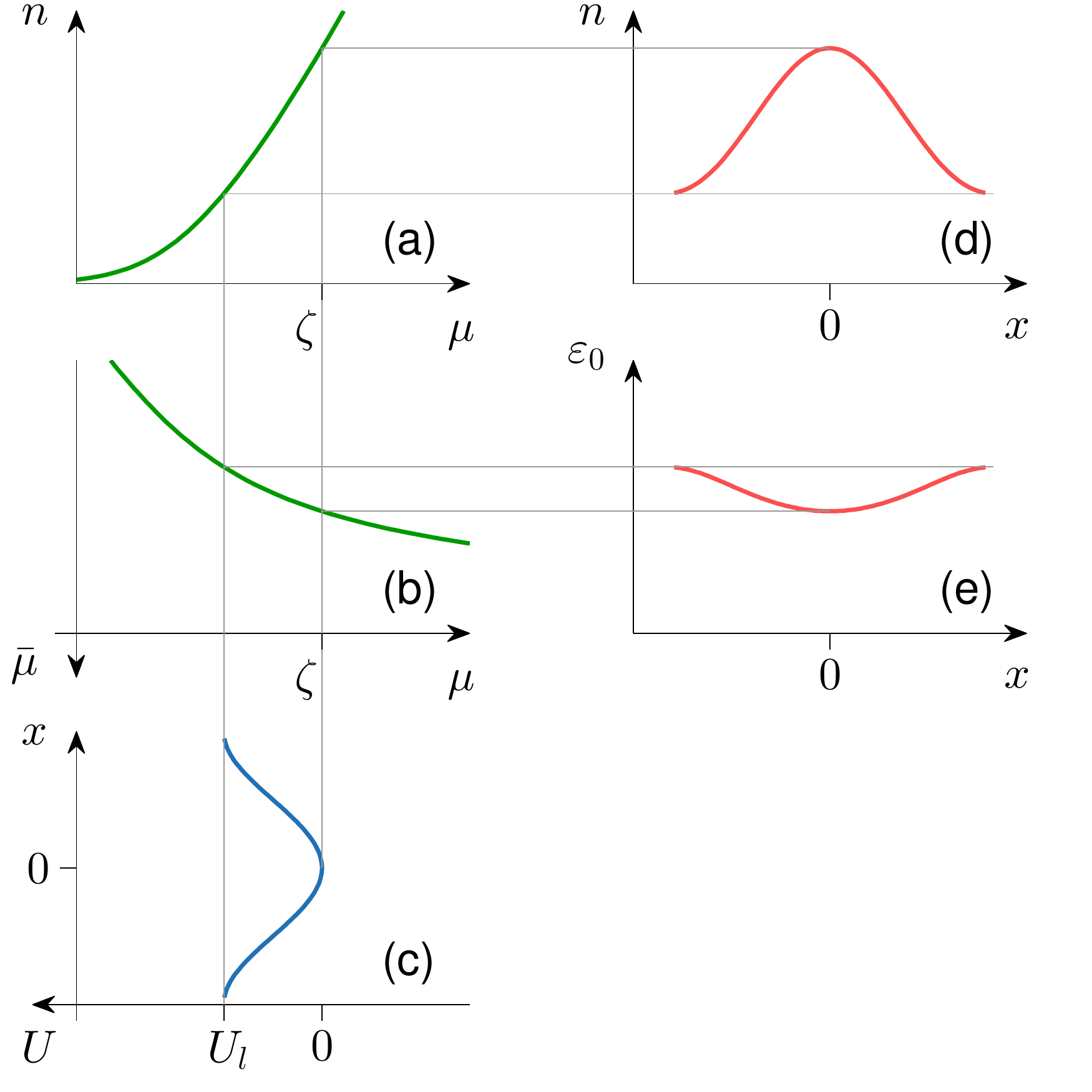}
\caption{Illustration of the relations among spatial modulations of the basic quantities in the problem. (a, b) Schematic reproductions of Fig.~\ref{fig:mu_and_n}(a, b) for one particular $\gamma$. (c, d, e) Variations of the lattice potential $U(x)$, exciton density $n$, and the exciton band edge $\varepsilon_0$ over one lattice period. Note the reversed axes direction in (b) and (c).
}
\label{fig:modulation_sketch}
\end{figure}

Let us examine the effect of the external potential $U(x)$ [Eq.~\eqref{eqn:E_0}] on the exciton normal fluid. If $U(x)$ does not vary too rapidly, then the semiclassical Thomas-Fermi approximation (TFA) is valid. Within the TFA the self-energy is considered to be position-dependent, $\Sigma(x) = n(x) t + U(x)$. It plays the role of the \textit{effective} potential acting on the excitons. The renormalized band edge $\varepsilon_0(x)$ is also position-dependent, tracking $\Sigma(x)$:
\begin{equation}
\varepsilon_0(x) = \bar\varepsilon_0 +
\Sigma(x) = \bar\varepsilon_0 + n(x) t + U(x)\,.
\label{eqn:A_from_A0_II}
\end{equation}
The local chemical potential $\mu(x)$ is defined such that Eq.~\eqref{eqn:n_from_mu} is unaltered. Finally, the occupation factors in Eqs.~\eqref{eqn:I_II}, \eqref{eqn:n_from_nu_bar}, \textit{etc.}, change from $f(\omega - \mu)$ to $f(\omega - \zeta)$,
where
\begin{equation}
\zeta = \mu(x) + U(x)
\label{eqn:zeta}
\end{equation}
is the electrochemical potential. One should not confuse $\zeta$, which is $x$-independent in equilibrium, with the spatially varying bare $\bar\mu(x)$ and renormalized $\mu(x)$ chemical potentials. By examining the last two we can analyze the screening of the external potential by the IXs. Indeed, from the above formulas we deduce
\begin{align}
\varepsilon_0(x) &= (\bar\varepsilon_0 + \zeta) - \bar\mu(x),
\label{eqn:varepsilon_0_from_mu_0}\\
\mu(x) &= \zeta - U(x)\,.
\label{eqn:mu_from_U}
\end{align}
Therefore, the plot of $\varepsilon_0$ \textit{vs.} $U$ can be obtained from the plot of $\bar\mu$ \textit{vs.} $\mu$ by inversion and a shift of the axes. This construction is illustrated in Fig.~\ref{fig:modulation_sketch}(b) and (e). The derivative
\begin{equation}
\begin{split}
\epsilon_* &\equiv
\frac{\partial U}{\partial \Sigma}
= \frac{\partial \mu}{\partial \bar\mu}
= 1 + t\, \frac{\partial n}{\partial \bar\mu}
\\
&= 1 + g \gamma \left(e^{n / n_d} - 1\right)\\
&= 1 + g \gamma \left[
\exp\left(\frac{\varepsilon_0 - \bar\varepsilon_0}{g \gamma T}\right) - 1
\right]\\
\end{split}
\label{eqn:epsilon_*}
\end{equation}
can be regarded as an effective dielectric constant. It describes how much variations of the effective potential $\Sigma(x)$ are reduced compared to those of the external one, $U(x)$. According to Eq.~\eqref{eqn:epsilon_*}, at fixed $n$ stronger interaction (higher $\gamma$) leads to stronger screening (larger $\epsilon_*$). Conversely, at fixed $\varepsilon_0$, higher $\gamma$ result in lower $\epsilon_*$, eventually approaching the interaction-independent limit
\begin{equation}
\epsilon_* = 1 + \frac{\varepsilon_0 - \bar\varepsilon_0}{T}\,.
\label{eqn:epsilon_*_limit}
\end{equation}
If the fixed parameter is $\gamma$, the screening evolves from negligible ($\epsilon_* \to 1$) to perfect ($\epsilon_* \to \infty$) as either $n$, $\varepsilon_0$, or $\mu$ increases. The last case is illustrated by Fig.~\ref{fig:mu_and_n}(c).

Small changes in $n$ and $U$, which are related by
\begin{equation}
\delta n(x) = -\delta U(x)\,
\frac{1 - \epsilon_*^{-1}}{t}\,,
\label{eqn:delta_n}
\end{equation}
are anti-correlated: maxima of $U(x)$ coincide with minima of $n(x)$, see Fig.~\ref{fig:modulation_sketch}(d). In the strong screening case, the density $n(x)$ is given, in the first approximation, by
\begin{equation}
n(x) \approx \frac{\zeta - U(x)}{t}\,
 \theta\bigl(\zeta - U(x)\bigr)\,,
\quad
\epsilon_*^{-1} \ll 1\,.
\label{eqn:n_strong_screening}
\end{equation}
Thus, the density profile is a mirror reflection of the external potential, except for intervals where $U(x) > \zeta$ if they exist. In those ``depletion regions'' the density is almost zero (exponentially small). To obtain the accurate solution for $n(x)$, we combine Eqs.~\eqref{eqn:n_from_mu} and \eqref{eqn:zeta} into
\begin{equation}
\exp\left(\frac{\bar\mu -\bar\varepsilon_0}{T}\right)
 + \exp\left(\frac{U - \zeta}{g \gamma T} + \frac{\bar\mu -\bar\varepsilon_0}{g\gamma T}\right) = 1\,,
\label{eqn:alpha_equation}
\end{equation}
which can be solved numerically~\cite{MATLAB} for $\bar\mu$ at given $\zeta$ and $U$, and then $\varepsilon_0$ and $n$ can be found. An example is shown in Fig.~\ref{fig:simulated_profiles}. The profile of $\varepsilon_0(x)$ [Fig.~\ref{fig:simulated_profiles}(a)] is seen to have a relatively flat minima, where screening is stronger and relatively sharp maxima, where screening is weaker. The screening is overall very efficient as the modulation of $\varepsilon_0(x)$ is almost an order of magnitude smaller than that of the external potential $U(x)$. Since $\zeta > U_l$ in this example, no depletion regions exist and the profile of $n(x)$ is nearly sinusoidal  [Fig.~\ref{fig:simulated_profiles}(b)].

\begin{figure}[t]
\centering
\includegraphics[width=3.2in]{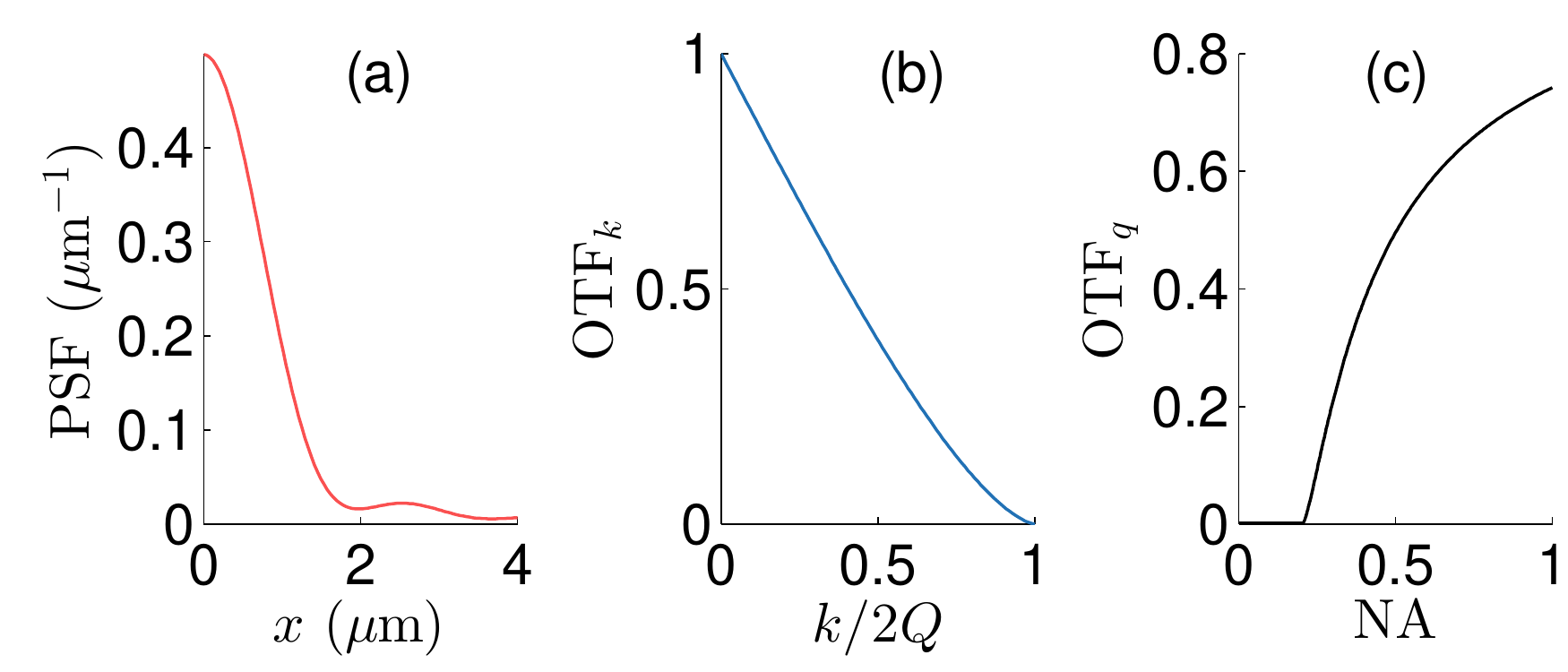}
\caption{Optical resolution functions.
(a) Point-spread function for $\mathrm{NA} = 0.24$. (b) Optical transfer function. Parameter $Q$ [Eq.~\eqref{eqn:Q}] is the largest momentum $k$ admitted by the optical imaging system.
(c) Optical transfer function as a function of the numerical aperture for the principal Fourier harmonic of momentum
$q = 2\pi / \Lambda$, $\Lambda = 2\unit{\mu\mathrm{m}}$.
}
\label{fig:OTF}
\end{figure}

\begin{figure}[b]
\centering
\includegraphics[width=3.2in]{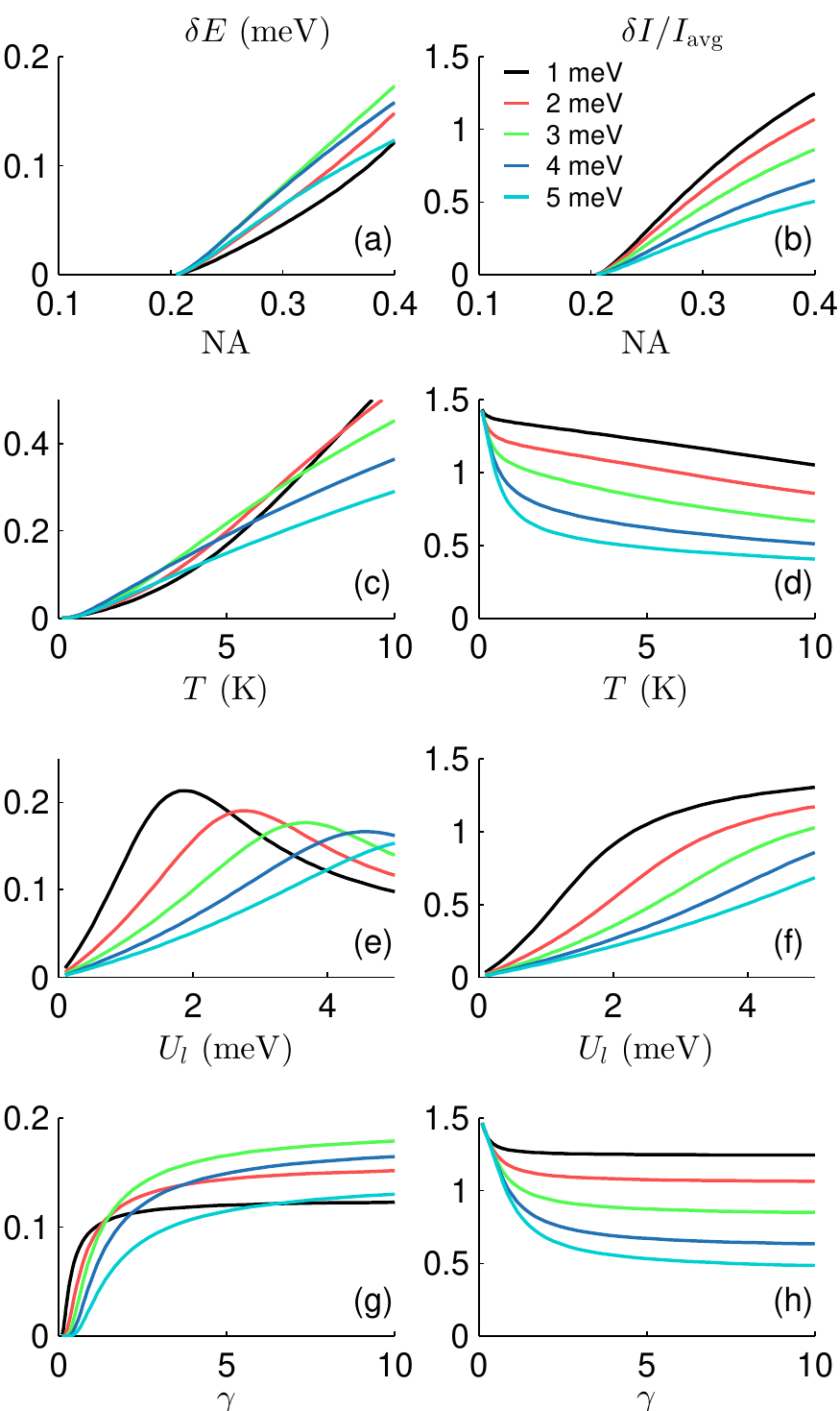}
\caption{(Color online) Parameter dependence of the PL energy $\delta E$ and the normalized intensity $\delta I / I_\mathrm{avg}$ modulations.
(a,b) $\delta E$ and $\delta I / I_\mathrm{avg}$ as functions of $\mathrm{NA}$ for $T = 4.2\unit{K}$, $U_l = 4\unit{meV}$, $\gamma = 7$.
(c,d) $\delta E$ and $\delta I / I_\mathrm{avg}$ as functions of temperature for $\mathrm{NA} = 0.4$, $U_l = 4\unit{meV}$, $\gamma = 7$.
(e,f) $\delta E$ and $\delta I / I_\mathrm{avg}$ as functions of the lattice depth for $T = 4.2\unit{K}$, $\mathrm{NA} = 0.4$, $\gamma = 7$.
(g,h) $\delta E$ and $\delta I / I_\mathrm{avg}$ as functions of the correlation parameter $\gamma$ for $T = 4.2\unit{K}$, $\mathrm{NA} = 0.4$, $U_l = 4\unit{meV}$.
The average ``blue shift'' $\Delta E_{\mathrm{avg}}$ for each color is indicated in the legend of panel (b).
}
\label{fig:parameter_dependence}
\end{figure}

Consider now the modulation of the PL spectrum $\tilde{I} = \tilde{I}(x, \omega)$. Combining Eqs.~\eqref{eqn:I_II} and ~\eqref{eqn:A_from_A0_II} gives
\begin{equation}
\tilde{I}(x, \omega) \mathop{=}\limits^{\displaystyle{?}}
 B f(\omega - \zeta) \bar{A}_0\bigl(\omega + \bar\varepsilon_0 - \varepsilon_0(x)\bigr).
\label{eqn:I_III}
\end{equation}
In order to account for the optical resolution, the convolution with the point-spread function (PSF) of the optical system must be added. For an ideal circular lens it is given by~\cite{Goodman2005ifo} $\mathrm{PSF}(x) = \mathrm{H}_1(2 Q x) / (\pi Q x^2)$,
where
\begin{equation}
Q = \mathrm{NA} \times k_0
\label{eqn:Q}
\end{equation}
is the largest momentum admitted by the lens, $\mathrm{NA} < 1$ is the numerical aperture,
and
\begin{equation}
k_0 = \omega\, /\, \hbar c
\label{eqn:k_0}
\end{equation}
is the photon momentum in vacuum. The plot of the PSF for $\mathrm{NA} = 0.24$, which is typical of our experimental setup, is shown in Fig.~\ref{fig:OTF}(a). It implies that the characteristic width of the PSF is comparable to the lattice period $\Lambda = 2 \pi / q = 2\,\mu\mathrm{m}$, so that accounting for the optical resolution effects is important. Having to evaluate the Struve function~\cite{Olver2010ndl} $\mathrm{H}_1(z)$ makes working with the PSF inconvenient. Instead, we can write the desired convolution as the Fourier series,
\begin{equation}
\tilde{I}(x, \omega) = \sum\limits_{k} \tilde{I}_k(\omega) \cos k x\,,
\label{eqn:I_FT}
\end{equation}
where $k = 0, q, 2 q, 3q, \ldots$ are the Fourier momenta and
\begin{equation}
\begin{split}
\tilde{I}_k(\omega) &= (2 - \delta_{k,0})B\, \mathrm{OTF}_k\, f(\omega - \zeta)
\\
&\times
\left\langle
\bar{A}_0\bigl(\omega + \bar\varepsilon_0 - \varepsilon_0(x)\bigr) \cos k x \right\rangle
\end{split}
\label{eqn:I_q}
\end{equation}
are the Fourier amplitudes. Neglecting the broadening of the spectral function once again, $\bar{A}_0(\omega) \to 2\pi \delta(\omega - \bar\varepsilon_0)$, the corresponding spectral moments are
\begin{equation}
\begin{split}
\tilde{I}_k^{(m)} &= (2 - \delta_{k,0}) B\, \mathrm{OTF}_k\\
&\times \left\langle
f\bigl(\varepsilon_0(x) - \zeta\bigr) [\varepsilon_0(x)]^m
\cos k x \right\rangle.
\end{split}
\label{eqn:I_m_q}
\end{equation}
Here and below $\langle z(x) \rangle \equiv \Lambda^{-1} \int_0^\Lambda z(x) d x$ is the average of a given function $z(x)$ over a lattice period. The optical transfer function~\cite{Goodman2005ifo} $\mathrm{OTF}_k$ is the Fourier transform of the PSF:
\begin{align}
\mathrm{OTF}_k &= \int
\frac{d^2 k^\prime}{\pi Q^2}\, \theta(Q - |\mathbf{k}^\prime|)\,
 \theta(Q - |\mathbf{k}^\prime - k \hat{\mathbf{x}}|)
\label{eqn:OTF_q}\\
 &= \frac{2}{\pi}
 \left[\arccos \left(\frac{k}{2Q}\right)
 - \frac{k}{2Q}\, \sqrt{1 - \frac{k^2}{4 Q^2}}\ \right]
\label{eqn:OTF_q_II}
\end{align}
for $k < 2 Q$.
Note the normalization: $\mathrm{OTF}_0 = 1$. At $k \geq 2 Q$, $\mathrm{OTF}_k$ is defined to be zero, which means that such harmonics are not resolved. The plot of $\mathrm{OTF}_k$ is shown in Fig.~\ref{fig:OTF}(b). The minimal numerical aperture necessary to observe at least the principal harmonic $k = q$ is therefore
\begin{equation}
\mathrm{NA}_{\mathrm{min}}  = \frac{q}{2 k_0}\,.
\label{eqn:NA_min}
\end{equation}
In our experiments (Sec.~\ref{sec:Experiments}) we are quite close to this lower limit, with $\mathrm{NA} \approx 0.24$ and
$\mathrm{OTF}_q \approx 0.07$, see Fig.~\ref{fig:OTF}(c).
This once again confirms that accounting for the optical resolution effects is important. Under the condition $1 < \mathrm{NA} / \mathrm{NA}_{\mathrm{min}} < 2$, which is satisfied in the experiment, all higher harmonics of spatial modulation are unobservable. Hence, the spectral moments $I^{(m)}(x)$ have the sinusoidal form, e.g.,
\begin{align}
I(x) &= I_{\mathrm{avg}}
 + \frac{\delta I}{2} \cos q x\,,
\label{eqn:I_harmonic}\\
I_{\mathrm{avg}} &= \tilde{I}_0^{(0)}\,,
\quad
\delta I = 2\tilde{I}_q^{(0)}\,,
\label{eqn:I_avg_def}\\
\frac{\delta I}{I_{\mathrm{avg}}}
 &= 4\, \mathrm{OTF}_q \,
  \dfrac{\left\langle
         f\bigl(E(0) - \bar\mu(x)\bigr) \cos q x
         \right\rangle}
        {\left\langle
         f\bigl(E(0) - \bar\mu(x)\bigr)
         \right\rangle}\,,
\label{eqn:deltaI_def}
\end{align}
Because of smallness of $\mathrm{OTF}_q$, the local PL energy is also approximately sinusoidal:
\begin{equation}
E(x) = \frac{\tilde{I}_0^{(1)} + \tilde{I}_q^{(1)}\cos q x}{\tilde{I}_0^{(0)} + \tilde{I}_q^{(0)}\cos q x}
 \approx E(0) + \Delta E_{\mathrm{avg}}
 - \frac{\delta E}{2} \cos q x\,.
\label{eqn:E_harmonic}
\end{equation}
Since $U(x)$ is correlated positively with $E(x)$ but negatively with $I(x)$, the modulation amplitudes $\delta I$ and $\delta E$ are positive. Once the electrochemical potential $\zeta$ is known, these amplitudes can be calculated numerically using $\bar\mu(x)$ found by solving Eq.~\eqref{eqn:alpha_equation}. We carried out such calculation for several sets of representative parameters, see below. In each set we fixed $\Delta E_{\mathrm{avg}}$, the directly measurable quantity. The corresponding $\zeta$ was determined by the standard root-searching algorithms.~\cite{MATLAB} The results are presented in Fig.~\ref{fig:parameter_dependence}. As one can see, the dependences of $\delta I$ and $\delta E$ on $\mathrm{NA}$ exhibit a threshold at $\mathrm{NA}_\mathrm{min}$ and an approximately linear growth thereafter [Figs.~\ref{fig:parameter_dependence}(a) and (b)]. These trends are inherited from the $\mathrm{OTF}_q$ (Fig.~\ref{fig:OTF}). Figures~\ref{fig:parameter_dependence}(e) and (f) show that $\delta E$ and $\delta I$ scale linearly with the lattice potential depth $U_l$ when it is small enough. The analytical expressions in this linear-response regime,
\begin{equation}
\Delta E_{\mathrm{avg}} \ll g \gamma T\,,
\quad
U_l \ll T + \Delta E_{\mathrm{avg}}\,,
\label{eqn:linear_regime}
\end{equation}
are as follows:
\begin{align}
\delta E \simeq
\mathrm{OTF}_q \, \frac{U_l T}{T + \Delta E_{\mathrm{avg}}}\,,
\quad
\frac{\delta I}{I_{\mathrm{avg}}}
\simeq
 \frac{\delta E}{T}
\label{eqn:deltaI}\,.
\end{align}
In this regime the modulation amplitudes are roughly independent of $\gamma$; however, they depend on the exciton temperature $T$ and the ratio of measured ${\delta I} / {I_{\mathrm{avg}}}$ and $\delta E$ can be used to determine it.

As $U_l$ increases and reaches $U_l \sim T + \Delta E_{\mathrm{avg}}$, the linear-response formulas cease to be valid. The density profile develops depletion regions around the maxima of $U(x)$, e.g., $x_0 < x < \Lambda - x_0$ where $\cos q x_0 = 1 - (2 \zeta / U_l)$. Outside the depletion regions, for example, at $|x| < x_0$, the density profile can be approximated by a vertically shifted cosine function: $n(x) \propto \cos q x - \cos q x_0$. The local energy $\varepsilon_0(x)$ at $|x| < x_0$ is close to $\zeta$ and no longer tracks $U_l$. For this reason, the energy modulation $\delta E$ goes through a maximum and then decreases. Concomitantly, ${\delta I} / {I_{\mathrm{avg}}}$ levels up at a plateau, see Figs.~\ref{fig:parameter_dependence}(e) and (f). Under the described conditions,
\begin{equation}
\frac{\delta I}{I_{\mathrm{avg}}} \simeq
\mathrm{OTF}_q\,
\frac{2 q x_0 - \sin 2 q x_0}{\sin q x_0 - q x_0 \cos q x_0}\,,
\label{eqn:nonlinear_regime}
\end{equation}
which is a function of the ratio $\zeta / U_0$ and independent of $T$ or $\gamma$. The maximum possible ${\delta I} / {I_{\mathrm{avg}}} = 4\, \mathrm{OTF}_q$, which is approximately $1.5$ for $\mathrm{NA} = 0.4$, is reached in the limit of $U_l \to \infty$ or $T \to 0$ or $\gamma \to 0$, see Figs.~\ref{fig:parameter_dependence}(f), (d), and (h), respectively. The energy modulation $\delta E$ vanishes in each of these three limits, see Figs.~\ref{fig:parameter_dependence}(c) and (g). We conclude that the dependence of ${\delta I} / {I_{\mathrm{avg}}}$ on the interaction parameter $\gamma$ is weak at both small and large $U_l$. However, this dependence is reasonably strong at the crossover point $U_l \sim T + \Delta E_{\mathrm{avg}}$ from the linear to nonlinear screening regime, see Fig.~\ref{fig:parameter_dependence}(h). This is also the ``sweet spot'' in terms of the $\gamma$-dependence of $\delta E$, see Fig.~\ref{fig:parameter_dependence}(g). Similarly, it appears that the intermediate temperature range $T \sim U_l$ offers the best conditions for estimating $\gamma$ from the measured $\delta E$ and $\delta I$.
Below in Sec.~\ref{sec:Experiments}, we present an experimental test of this estimation method.

\section{Experiments}
\label{sec:Experiments}

\begin{figure}
\centering
\includegraphics[width=7.5cm]{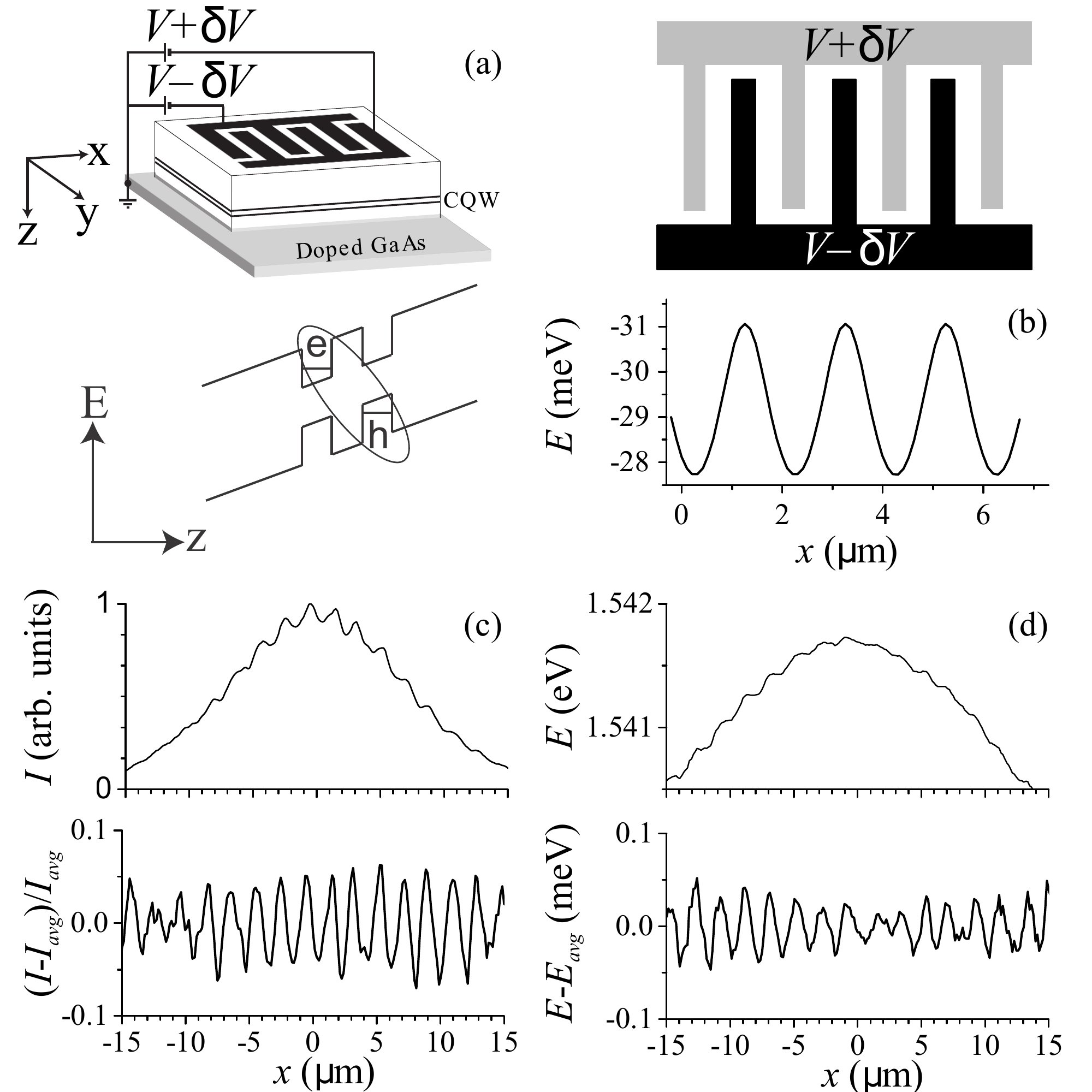}
\caption{Sample structure and experimental results for indirect excitons in electrostatic lattices. (a) Schematic of the CQW sample (top) and its band diagram (bottom). GaAs quantum wells are positioned within the insulating Al\sub{0.33}Ga\sub{0.67}As layer (white) surrounded by conducting $n^+$ GaAs layers (gray) at the bottom and interdigitated electrodes on the sample surface at the top. Labels ``e'' and ``h'' indicate an electron an a hole bound into an indirect exciton. (b) Electrode schematic (top) and the lattice potential profile (bottom) for $V_0 = 3\unit{V}$ and $\delta V = 0.5\unit{V}$ from electrostatic simulations. (c) Top: measured PL intensity profile of indirect excitons. Bottom: the same data after subtraction of a smooth background and normalization, showing the periodic modulation more clearly. (d) Top: PL energy profile of indirect excitons. Bottom: the same data after subtraction of a smooth background. Experimental parameters for (c) and (d) are: bath temperature $T_\mathrm{bath} = 1\unit{K}$, lattice depth $U_l = 4.8\unit{meV}$, photoexcitation power $P = 8\unit{\mu}\mathrm{W}$.
}
\label{fig:Experiment}
\end{figure}

Our experiments were performed on CQW structure grown by molecular-beam epitaxy. Two $8\unit{nm}$-wide GaAs quantum wells separated by a $4\unit{nm}$-thick Al$_{0.33}$Ga$_{0.67}$As barrier were positioned $100\unit{nm}$ above the $n^+$-type GaAs layer within an undoped $1\,\mu\mathrm{m}$-thick Al$_{0.33}$Ga$_{0.67}$As layer. The conducting $n^+$ GaAs layer at the bottom served as a ground electrode, see Fig.~\ref{fig:Experiment}(a). Semitransparent interdigitated top electrodes were fabricated by magnetron sputtering a $90\unit{nm}$ indium tin oxide layer. These electrodes generate a laterally modulated electric field perpendicular to the quantum well plane, which couples to the static dipole momentum $e d$ of the IXs.~\cite{Zimmermann1998sre, Remeika2009ldt, Winbow2011ece, Remeika2012tde} The average potential was fixed by the average voltage $V_0 = 3\unit{V}$ and the lattice depth $U_l$ was controlled by the voltage difference $\delta V$, Fig.~\ref{fig:Experiment}(b). The measurements were performed in an optical dilution refrigerator. The refrigerator had high stability with the vibration amplitude well below the lattice period $\Lambda = 2.0\unit{\mu}\mathrm{m}$. The IXs were generated by a cw $633\unit{nm}$ laser focused to an excitation spot of diameter $\sim 10\unit{\mu}\mathrm{m}$. The exciton density was controlled by the laser excitation power.

\begin{figure}
\centering
\includegraphics[width=7.5cm]{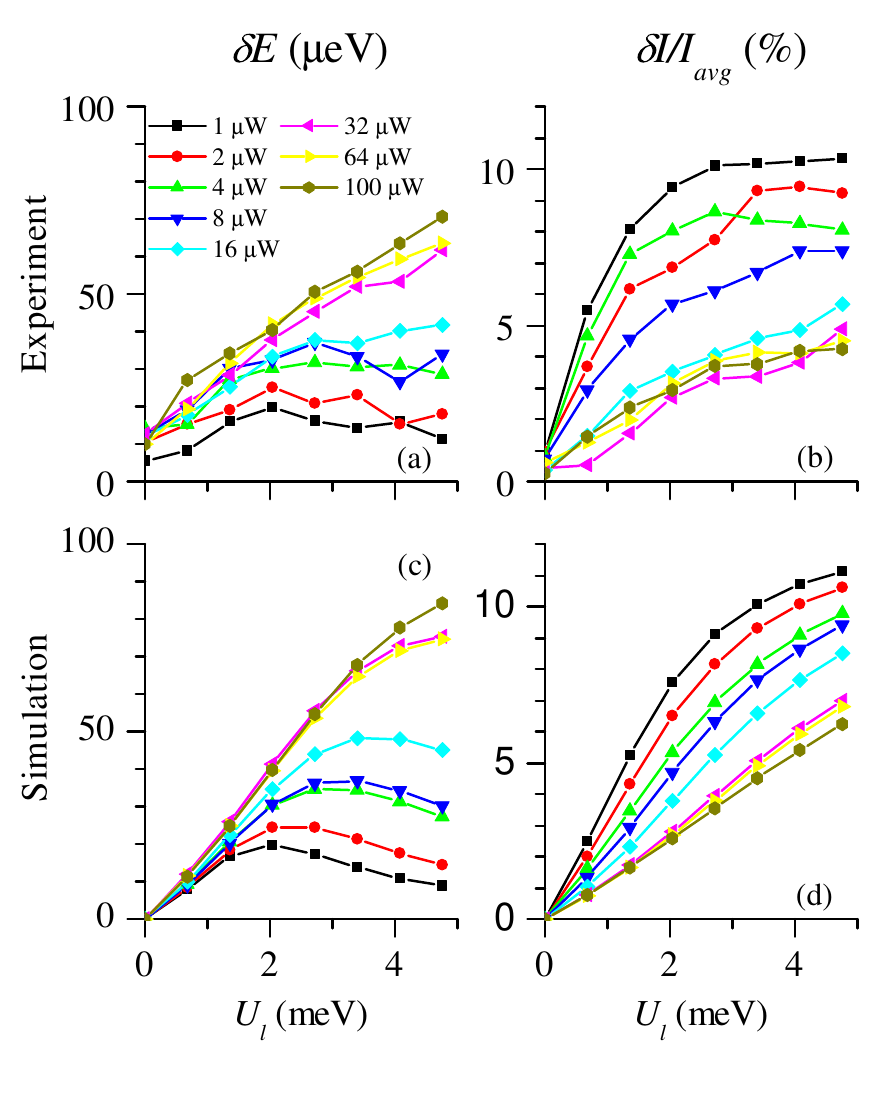}
\caption{(Color online) Measured and simulated modulation of energy and emission intensity of indirect excitons in a lattice.
(a,b) Measured energy (a) and emission intensity (b) modulation for indirect excitons as a function of the lattice depth for different laser excitation powers (shown in the legend). $T_\mathrm{bath} = 1\unit{K}$. (c,d) Simulated energy (c) and emission intensity (d) modulation. For each curve, the electrochemical potential is selected to correspond to the measured exciton interaction energy at the excitation power in (a,b). $\gamma$ and $T$ are optimized to obtain best fit to the experimental energy and intensity modulation curves.}
\label{fig:modulation_fits}
\end{figure}

Examples of the PL energy and intensity profiles measured for IXs in the lattice are presented in Fig.~\ref{fig:Experiment}.
Figures~\ref{fig:modulation_fits}(a) and (b) depict $\delta E$ and  $\delta I / I_\mathrm{avg}$ as a function of $U_l$ for several different excitation powers. The same quantities computed using the model of Sec.~\ref{sec:Model} are shown in Fig.~\ref{fig:modulation_fits}(c) and (d). For each simulated curve, the electrochemical potential was adjusted to match the average PL energy in the excitation spot center. The parameter $\gamma$ and IX temperature were optimized to obtain best fit to the experimental data for both energy and intensity modulation curves (Fig.~\ref{fig:modulation_fits}a,b).

\section{Discussion}
\label{sec:Discussion}

IX gas in CQW is a model system for studying dipolar matter because many of its basic physical parameters can be controlled experimentally. For example, the range of accessible $n$ can span several decades. This opens an opportunity to experimentally test various theoretical predictions regarding how many-body correlations of dipolar bosons evolve as a function of the particle density. Such a comparison with theory requires the development of a method for accurately determining the exciton density $n$ in absolute units that remains a challenging problem. For example, estimation of $n$ from time-integrated exciton emission suffers from large uncertainties.~\cite{OHara1999snr} The method for determining exciton density by measuring the Landau level filling factors of electrons and holes\cite{Butov1991edn} is accurate, however it requires high magnetic fields. In that regime the exciton properties are strongly modified compared to the zero-field case. In fact, in the limit of high magnetic field the interaction between spatially direct excitons vanishes~\cite{Lerner1981tde} that was verified experimentally.~\cite{Butov1992edt} The recently proposed technique of remote electrostatic sensing is promising but challenging to implement.~\cite{Cohen2011rdi} Compared to all of the above, the lattice-based method proposed in our earlier work~\cite{Remeika2009ldt} and developed further in the present article appears to be an attractive alternative.

\begin{figure}
\centering
\includegraphics[width=5.5cm]{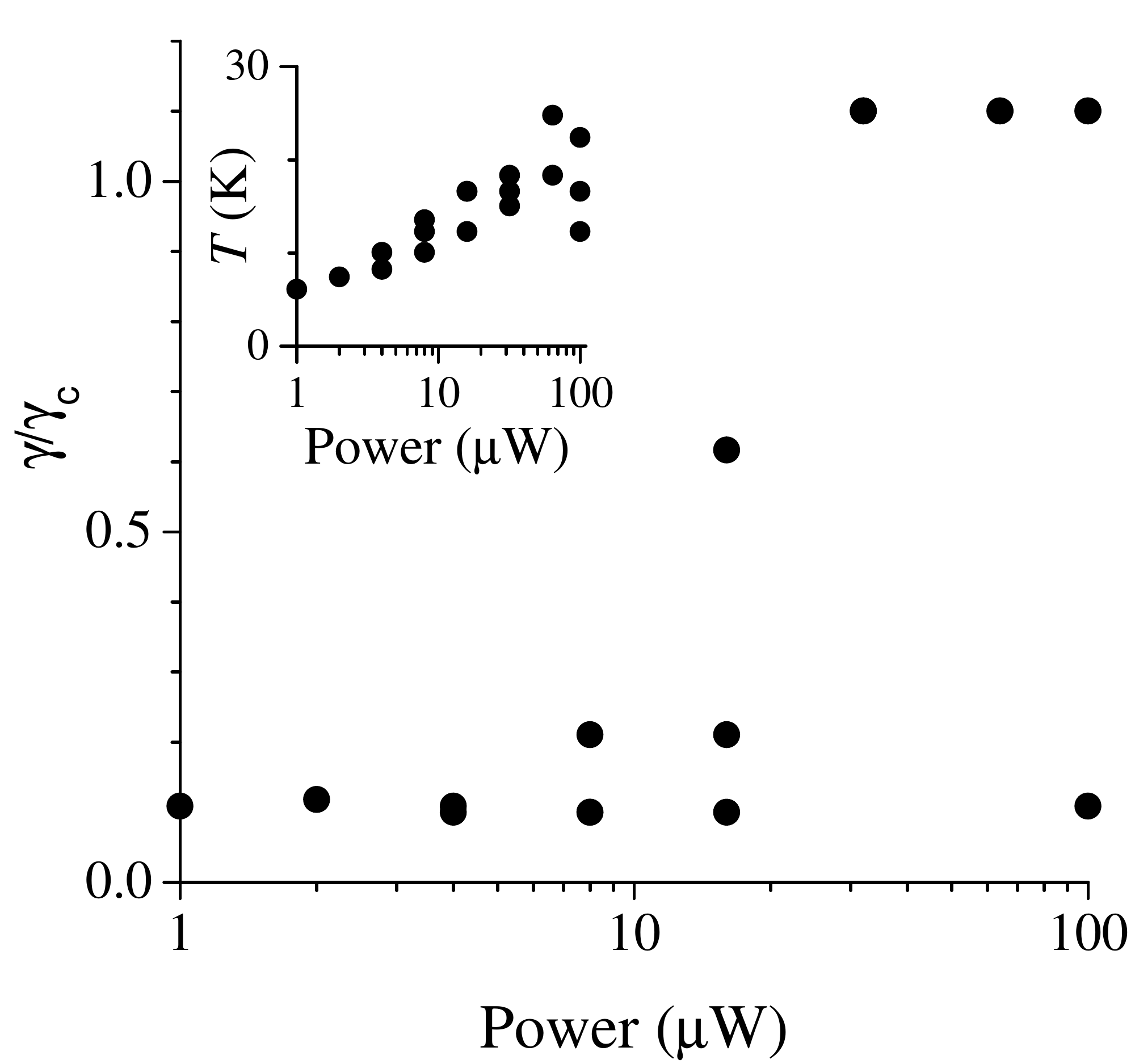}
\caption{(Color online) Estimated IX correlation parameter $\gamma$ obtained from fitting the experimental data for both energy and intensity modulation curves in Figs.~\ref{fig:modulation_fits}(a) and (b). The values of $\gamma$ are normalized to $\gamma_c = 7$ [Eq.~\eqref{eqn:gamma_plate}].
The inset shows the temperature values for these fits.
}
\label{fig:GammaFit}
\end{figure}

The qualitative agreement of experimental and simulated ${\delta}E$ (Fig.~\ref{fig:modulation_fits}a,c) and ${\delta}I/I$ (Fig.~\ref{fig:modulation_fits}b,d) constitutes a proof-of-principle demonstration of the method. The accuracy of the data is still low as seen by the large data scattering in Fig.~\ref{fig:GammaFit}. The presented theory and experiment provide a guide for improvement. Since both modulation amplitudes increase with $\mathrm{NA}$ (Fig.~\ref{fig:parameter_dependence}a,b), it is advantageous to use an objective with a highest possible $\mathrm{NA}$. Larger period of the lattice potential may also be helpful. Although the fitted values of the exciton temperature $T$ (Fig.~\ref{fig:GammaFit}, inset) are in agreement with those from earlier studies,~\cite{Ivanov2006oir, Hammack2006tce, Hammack2007kie, Hammack2009kir, Butov2001ssi, Remeika2013pfe}
the accuracy of the method can be further improved if $T$ is known independently. This can be achieved by measuring the exciton PL after a pulsed excitation in a time interval of the order of a few nanoseconds after the excitation pulse. Under such conditions, the IXs cool down and their temperature approaches the bath temperature.~\cite{Hammack2007kie, Hammack2009kir} Using a defocused laser excitation spot to create an IX cloud with a more uniform density may also be beneficial. Such improvements of the proposed method is a subject for future experiments.

On the theory side, better understanding of the disorder effects on the exciton PL spectrum is necessary. This may be particularly important for correctly interpreting the data points for low powers [such as $1\,\mu\mathrm{W}$ and $2\,\mu\mathrm{W}$, black squares and red dots in Fig.~\ref{fig:modulation_fits}(a) and (b)]. Here we confine ourselves to the following brief discussion. As mentioned in Sec.~\ref{sec:Introduction}, efficient loading of excitons into the lattice requires working with systems where the blue shift $\Delta E_\mathrm{avg}$ is higher than the characteristic energy scale of the disorder. In this case the interaction of excitons with random potential of disorder can be treated perturbatively. Thus, we can use the linear-response (effective) dielectric function $\epsilon_*(k)$ to determine how the exciton gas screens this random potential.
The zero-$k$ limit of $\epsilon_*(k)$ is given by Eqs.~\eqref{eqn:epsilon_*} and \eqref{eqn:epsilon_*_limit}, while its large-$k$ behavior can be shown to be of the form~\cite{Kachintsev1994dfc} $\epsilon_*(k) = 1 + (8\pi\gamma n / k^2)$. Since $\epsilon_*$ depends on $n$, so does the disorder-induced self-energy correction $\Sigma_{d}$. Assuming the bare random potential is a white noise of bare strength $\bar{v}_d$, we use the first Born approximation to get
\begin{equation}
\begin{split}
\Sigma_{d}(\omega) &= \int \frac{d^2 k}{(2\pi)^2}
\frac{\bar{v}_d}{\omega - \bar\varepsilon_{\mathbf{k}}}
\,
\left[\frac{1}{\epsilon_*^2(k)} - 1\right]
\\
&\simeq
-i \Gamma + \frac{\Gamma}{\pi}\, \ln \left(\frac{2 e n t}{\omega - \bar\varepsilon_0}\right)
\,,
\quad
\Gamma \equiv \bar\nu_1 v_d\,.
\end{split}
\label{eqn:Sigma_BA}
\end{equation}
The next improvement is the self-consistent Born approximation,~\cite{Thouless1978} which is obtained replacing $\bar\varepsilon_0$ in the logarithm with its renormalized value $\varepsilon_0$. For $T \gg \Gamma$, the energies $\omega$ relevant for the PL are $\omega - \varepsilon_0 \sim T$. The corresponding real part of the self-energy is
\begin{equation}
\Sigma_{d}^{\prime} \simeq \frac{\Gamma}{\pi}\,
\ln \left(\frac{2 e n t}{T}\right)\,.
\label{eqn:sigma_disorder}
\end{equation}
Using the definition~\eqref{eqn:gamma} of $\gamma$ and replacing $n t$ by $\Delta E_\mathrm{avg}$, we find that the random potential creates a fractional correction
\begin{equation}
\frac{\gamma_{d}}{\gamma} \simeq \frac{1}{\pi}\, \frac{\Gamma}{\Delta E_\mathrm{avg}}\,
\ln \left(2 e\, \frac{\Delta E_\mathrm{avg}}{T}\right)
\label{eqn:gamma_disorder}
\end{equation}
to the correlation parameter $\gamma$. For $\Delta E_\mathrm{avg} = 0.5$--$2\unit{meV}$ in Fig.~\ref{fig:modulation_fits}(a) and (b), this correction can be as high as $\sim 50\%$ assuming $\Gamma, T \sim 1\unit{meV}$ (Figs.~\ref{fig:phases} and \ref{fig:GammaFit}).

Finally, we wish to point out that our method of determining the correlation parameter of an interacting gas is quite general.
Similar techniques can be applied for a variety of other systems that can be subjected to lattice potentials and whose spatially modulated parameters can be measured. Examples include cold atomic clouds,~\cite{Greiner2002qpt, Chin2006esu} exciton polariton systems,~\cite{Lai2007czs, Bajoni2008plu, Cerda-Mendez2010pcd} and two-dimensional electron layers.~\cite{Singha2011tmh}

\acknowledgments

This work was supported by the DOE Office of Basic Energy Sciences under award DE-FG02-07ER46449.
MMF was supported by the University of California Office of the President.

%

%

\end{document}